\documentclass[12pt]{article}
\usepackage{epsfig}
\usepackage{amsfonts}
\usepackage{amscd}
\usepackage{latexsym}
\usepackage{amsmath,amssymb}
\usepackage{verbatim}
\usepackage{setspace}
\usepackage{color}
\usepackage{cite}

\usepackage[textheight=9in, textwidth=6.5in, letterpaper]{geometry}

\usepackage{hyperref}
\hypersetup{
    colorlinks,
    citecolor=black,
    filecolor=black,
    linkcolor=black,
    urlcolor=black
    linktoc=all
}

\numberwithin{equation}{section}
 
 \def\p{\partial}
 \def\bz{{\bar z}} 
  \def\bw{{\bar w}} 
  \def\ip{${\cal I}^+$}
\def\0{{(0)}}
\def\1{{(1)}}
\def\2{{(2)}}

\def\cg{{\mathcal G}}
\def\ch{{\mathcal H}_3}
\def\ci{{\mathcal I}}

\def\cs{{\mathcal S}} 
\def\cw{{\mathcal W}} 
 
\def\cst{${\cal C S}^2$}
\newcommand{\bea}{\begin{eqnarray}}
\newcommand{\eea}{\end{eqnarray}}
\newcommand{\be}{\begin{equation}}
\newcommand{\ee}{\end{equation}}
\newcommand{\ba}{\begin{align}}
\newcommand{\ea}{\end{align}}
\def\be{\begin{equation}}
\def\ee{\end{equation}}
\def\beq{\be\begin{array}{c}}
\def\eeq{\end{array}\ee} 
 
\newcommand{\ve}{\varepsilon}  
 
\def\btau{\bar{\tau}}
\def\bJ{\bar{J}}
\def\bS{\bar{S}}

\def\slc{SL$(2,\mathbb{C})$}
\def\gc{{\Gamma_{\text{cusp}}}}

   \makeatletter
  \let\over=\@@over \let\overwithdelims=\@@overwithdelims
  \let\atop=\@@atop \let\atopwithdelims=\@@atopwithdelims
  \let\above=\@@above \let\abovewithdelims=\@@abovewithdelims
\renewcommand\section{\@startsection {section}{1}{\z@}%
                                   {-3.5ex \@plus -1ex \@minus -.2ex}
                                   {2.3ex \@plus.2ex}%
                                   {\normalfont\large\bfseries}}

\renewcommand\subsection{\@startsection{subsection}{2}{\z@}%
                                     {-3.25ex\@plus -1ex \@minus -.2ex}%
                                     {1.5ex \@plus .2ex}%
                                     {\normalfont\bfseries}}

\begin{document}
\begin{titlepage}
\unitlength = 1mm

\ \\
\vskip 1cm
\begin{center}

{ \LARGE {\textsc{ Soft Factorization in QED from 2D Kac-Moody Symmetry}}}

\vspace{0.8cm}
Anjalika Nande, Monica Pate, and Andrew Strominger

\vspace{1cm}

{\it  Center for the Fundamental Laws of Nature, Harvard University,\\
Cambridge, MA 02138, USA}\\

\begin{abstract}

The soft factorization theorem for 4D abelian gauge theory states that 
the $\cs$-matrix factorizes into soft and hard  parts, with the universal soft part containing all soft and collinear poles. Similarly,  correlation functions on the sphere in a 
2D  CFT with a $U(1)$ Kac-Moody current algebra factorize into current algebra and non-current algebra factors, with the current algebra factor fully determined by its pole structure. In this paper, we show that these 4D and 2D factorizations are mathematically the same phenomena. The soft `tHooft-Wilson lines and soft photons are realized as a complexified 2D current algebra
on the celestial sphere at null infinity. The current algebra   level is determined  by the cusp anomalous dimension. The  associated complex $U(1)$ boson lives on a torus  whose modular parameter is $\tau =\frac {2\pi i }{e^2}+{\theta \over 2 \pi}$. The correlators of this 2D current algebra fully reproduce 
the known soft part of the 4D $\cs$-matrix, as well as a conjectured generalization involving magnetic charges.

  \end{abstract}

\vspace{1.0cm}

\end{center}

\end{titlepage}

\pagestyle{empty}
\pagestyle{plain}

\pagenumbering{arabic}

\tableofcontents

\section{Introduction}

The scattering amplitudes  of any 4D quantum field theory can be rewritten
in the form of correlation functions on the `celestial sphere' at null infinity, denoted \cst \cite{He:2015zea} . For massless particles one  simply represents asymptotic particle states by  an operator (carrying labels such as energy, spin or charge) at the point on \cst\ at which they enter or exit Minkowski space. Massive particles - the focus of this paper - are labelled by a point on the unit 3D hyperbola $\ch$ rather than \cst. However using the bulk-to-boundary propagator on $\ch$, these have a precise representation as smeared operators on \cst\ \cite{Campiglia:2015qka,Campiglia:2015lxa, Kapec:2015ena,Pasterski:2016qvg}. 

The utility of so rewriting the $\cs$-matrix devolves from the fact that the Lorentz group acts as the global conformal group \slc\  on \cst. This implies that these correlators share many properties with those of a 2D CFT on the sphere and are subject to many of the same constraints. This connection is strengthened by the observation \cite{Kapec:2014opa} that, once coupled to gravity, the global group is enhanced to the full infinite-dimensional local conformal group.

In this paper we further explore the connection between 4D scattering amplitudes in abelian gauge theory and 2D  CFT.  
It has already been noted \cite{Strominger:2013lka,He:2015zea} that soft photon insertions in the 4D $\cs$-matrix are equivalent to insertions of a $U(1)$ Kac-Moody current on \cst. 
2D CFTs with such a current enjoy  a well-known factorization property: every correlator on the sphere is a product of a current algebra correlator with 
a correlator of `stripped'  operators which decouple from the current algebra. It is natural to ask if this 2D factorization lifts to the full 4D $\cs$-matrix. Indeed we show this is precisely the content of the `soft factorization theorem' \cite{Bassetto:1984ik,Feige:2014,SchwartzQFT} in 4D abelian gauge theory. The latter states that the $\cs$-matrix factorizes into a hard and soft part, distinguished by the introduction of an IR scale. The leading soft $\cs$-matrix contains all soft and collinear poles. We show that  it can be efficiently reproduced by 
the 2D $U(1)$ current algebra with the level given by the cusp anomalous dimension. 

The literature on the abelian soft factor largely concerns electrically charged asymptotic particles. However when expressed as a 2D current algebra, the general case including magnetic charges \cite{Strominger:2015bla}
 is extremely natural and in some respects simpler than the pure electric case. 
The soft photon current  is complexified and is the gradient of 
a 2D scalar living on the torus whose modular parameter is $\tau =\frac {2\pi i }{e^2}+{\theta \over 2 \pi}$. This scalar can be identified as the Goldstone boson of the spontaneously broken large electric and magnetic gauge symmetries. We use this current algebra to derive the soft $\cs$-matrix when asymptotic particles carry magnetic as well as electric charges, and conjecture that it is non-perturbatively exact. 

This paper is organized as follows. Section 2 establishes our conventions and reviews the \cst\ smearing function for massive particles in terms of the bulk-to-boundary propagator on $\ch$. In section 3, the 2D current algebra description of soft electric Wilson lines is given. In section 4, the 2D currents are related to the gauge field at the boundaries of 
null infinity. Section 5 computes soft Wilson line expectation values and identifies the current algebra level with the cusp anomalous dimension. 
In section 6, magnetic Wilson (`tHooft) operators  are incorporated, duality transformation properties are studied and an identification of the complexified soft photon and Goldstone currents is proposed. In section 7, the  expectation value of general electromagnetically  charged Wilson lines with soft photon insertions is computed. 
Section 8 relates this formula  to the soft part of the $\cs$-matrix in abelian gauge theory.

\section{Preliminaries} \label{sec:largegaugesym}

	In this section, we establish notation and review the electric large gauge symmetry of abelian gauge theories.   We largely follow the conventions in the review \cite{Strominger:2017zoo}.
	In the past and future lightcones of the origin, we  adopt hyperbolic coordinates for which the Minkowski line element is 
	\be
		ds^2 = - d \tau^2  + \tau^2 \left (\frac{d\rho^2}{\rho^2} + \rho^2 dz d\bz\right)~.
	\ee 
	Surfaces of constant $\tau^2=-x^\mu x_\mu$ are the hyperbolic plane $\ch$, while $(z,\bz)$ parametrizes the celestial sphere denoted  \cst. 
These coordinates are related to the standard Cartesian coordinates  $x^\mu$ by 
	\be \label{eq:massivecoord}
		x^\mu = \frac{\tau  }{2 \rho } \left ( n^\mu  +  \rho^2 \hat q^\mu (z, \bz)\right)~, 
	\ee
	where $n^\mu$ and $q^\mu(z, \bz)$ are the null vectors
	\be
		\begin{split}
			n^\mu \equiv  (1, 0,0,-1)~, \quad \quad \quad 
			\hat q^\mu (z, \bz) \equiv \big(1+ z \bz, z+\bz, -i (z-\bz), 1-z \bz\big)~,
		\end{split}
	\ee
which obey
	\be
		n^\mu \hat q_\mu (z, \bz)  = -2~, \quad \quad \quad
		 \hat q^\mu (z, \bz) \hat q_\mu (w, \bw) = -2 |z-w|^2~.
	\ee
The region of real $\tau$  is the interior of the lightcone emanating from the origin. The region outside this lightcone corresponds to imaginary $(\tau,\rho)$ and a dS$_3$ slicing but will not be needed herein. These coordinates provide
	a resolution of future (past) timelike infinity $i^+$ ($i^-$) to a spacelike hyperboloid $\ch^+$ ($\ch^-$) that is obtained by taking the limit $ \tau \to \infty$ at fixed $(\rho,z, \bz)$ with $\rho>0$ ($\rho<0$). 
	The boundary of any $\ch$ with positive (negative)  $\rho $ is the $S^2$ intersection of the future (past) lightcone of the origin with null infinity $\ci$. 
Points at a given value of  $(z,\bz)$ on $\ci^-$ lie at antipodal angles to those at the same value on $\ci^+$. 
	 	
	We consider theories with a $U(1)$ gauge field $F = dA$ obeying the Maxwell equation 
	\be 
		d*F = e^2*j~.
	\ee   
	Such theories are invariant under electric gauge transformations of the form
	\be \label{eq:symtransf}
		A_\mu (x) \to A_\mu(x) +\p_\mu \ve (x)~,\quad \quad  \Psi_k (x)  \to e^{i {Q_k} \ve (x)} \Psi_k (x)~,
	\ee
	where $\ve \sim \ve + 2 \pi $ and $\Psi_k$ is a matter field of charge ${Q_k} \in   \mathbb{Z}$.   
	
	Gauge transformations that vanish at the boundary of Minkowski space generate redundant descriptions of the same physical state and hence are trivial.  We consider  non-trivial `large' gauge transformations  
	\cite{Strominger:2013lka,He:2014cra} which approach a $u$-independent  function \be\label{eb}  \ve (z,\bz)\equiv \ve (z,\bz, u = \tau/\rho , r = \tau \rho=\infty)\ee  at \ip\ and the boundary of $\ch^+$.
	In Lorenz gauge $\nabla^\mu A_\mu=\square  \ve=0$, the bulk gauge parameter that generates this large
	gauge symmetry is
	\be \label{eq:bulkeps}
		 \ve (x) = \int d^2 w ~\cg \big(x; w,\bw\big) \ve (w,\bw)~,
	\ee
	where
	\be \label{def:propagator}
		\cg  \big(x; w, \bw\big)= - \frac{1}{2 \pi } \frac{x^2 }{ \left [x \cdot \hat q(w, \bw)\right]^2}~.
	\ee
Evaluating $\cg$ in the coordinate system \eqref{eq:massivecoord}, one finds it is $\tau$-independent
	 \be
	  	 \cg  \big(\rho, z, \bz; w,\bw\big) = \frac{1}{2 \pi } \frac{\rho^2}{\left[\rho^2 |z-w|^2 +1\right]^2}~.
	  \ee
	 	  Moreover, near  the boundary of $\ch^+$ 
	  \be
	  	\lim_{\rho \to \infty} \cg  \big(\rho, z, \bz; w,\bw\big)  = \delta^2 (z-w)~.
	  \ee This immediately implies that  the bulk expression \eqref{eq:bulkeps} for  $ \ve$ in Lorenz gauge obeys \eqref{eb}. $\cg$ may be recognized as  the familiar bulk-to-boundary propagator for a massless scalar 	on  $\ch$ encountered in studies of AdS/CFT. 
Finally, 	it follows from the construction \eqref{eq:bulkeps} that  $\ve$  obeys 
	  	\be \label{eq:matchingcondition}
					  \ve (z,\bz ) \big|_{\ci^+} = \ve (z,\bz )\big |_{ \ci^-}~,
		\ee
and hence obeys the antipodal matching condition for the electric large gauge symmetries of QED \cite{He:2014cra,Campiglia:2015qka,Kapec:2015ena}.  
\section{Soft photon current algebra}

		\label{sec:softphotoncurrent} 
	
In this section we describe the soft photon insertions and soft Wilson lines that comprise the soft sector of the $\cs$-matrix. 
Any 4D $\cs$-matrix element involving  $n$ massless particles  may, by a change of notation,  be rewritten as  a 2D `correlation function' on \cst\ with $n$ operator insertions
	\be\label{cfn}
		\mathcal A_n  = \langle O_1 (z_1, \bz_1) \cdots  O_n (z_n, \bz_n)   \rangle~,
	\ee
 	 where each operator $O_k(z_k, \bz_k)$ creates or annihilates  a massless particle which pierces null infinity at 
	 \be 
	 	z_k={p^1_k+ip^2_k \over  E_k+p^3_k }~,
	\ee 
where we take 	 $E_k$  to be positive (negative) for outgoing (incoming) particles. 
	 $O_k$ in general carries other suppressed labels such as the spin or electric charge $Q_k$.  \slc\ Lorentz invariance implies global conformal invariance of the correlation functions \eqref{cfn}. 

In this paper, we seek to describe the coupling to the gauge field of timelike Wilson lines on asymptotic trajectories $x_k^\mu(s)=p_k^\mu s$ describing massive\footnote{Complications arise for massless charged particles at the loop level because the coupling constant runs in the deep IR.} charged particles. Massive particles in momentum eigenstates do not reach \ip\ and cannot be associated with local operators at unique points on \cst.\footnote{There is however an alternate basis of boost-eigenstate wave functions which are associated to local operators on \cst \cite{Pasterski:2016qvg}.  } Instead $p_k\over m_k$, which obeys ${p_k^2\over m_k^2}=-1$,  labels a point on the asymptotic hyperbola $\ch^+$ describing future timelike infinity \cite{Campiglia:2015qka,Pasterski:2016qvg}. The Wilson line for a charge $Q_k$ particle is described by a smeared operator of the form \cite{Campiglia:2015qka,Campiglia:2015lxa,Kapec:2015ena}
\be \label{eq:defdressing}
					\cw_{Q_k}(    p_k) \equiv \exp \left [iQ_k \int d^2 w~ \cg \big(p_k; w, \bw  \big) \phi^E(w,\bw ) \right]~, \quad \quad	\ee  
where the bulk-to-boundary propagator $\cg$ was given in \eqref{def:propagator}  and $\phi^E$   is a local operator (constructed from the boundary gauge field in the next section) on \cst\ which we refer to as the electric Goldstone boson.
The superscript $E$ distinguishes it from its magnetic counterpart  which will appear later.  Under a large gauge  transformation it transforms as a Goldstone mode 	\be\label{szw}
		\delta_\ve \phi^E(z,\bz ) = { \ve (z,\bz ) }~,
	\ee
	implying a periodicity 
	\be  \phi^E \sim \phi^E +  2 \pi ~.\ee
As shown in \cite{Campiglia:2015qka,Campiglia:2015lxa,Kapec:2015ena}
the smearing function  is uniquely determined by the symmetries at hand  to be the bulk-to-boundary propagator $\cg$.  Under a large gauge transformation one easily finds using  \eqref{szw}
\be \label{nr} \delta_\ve\cw_{Q_k}(    p_k) =   iQ_k  {\ve}(p_k)\cw_{Q_k}(    p_k)~, \quad \quad	\ee 
as required by \eqref{eq:symtransf}. 
It has the property that it approaches a delta-function as the particle is boosted to the speed of light so that the exponent of \eqref {eq:defdressing} becomes local.  As $E_k \to \infty$ at fixed mass,  
\be \label{ult}  
	\cw_{Q_k}(p_k)\to e^{iQ_k\phi^E(z_k, \bz_k )}\equiv \cw_{Q_k}(z_k, \bz_k )~, \quad~~~ \delta_\ve \cw_{Q_k}( z_k, \bz_k )  = iQ_k  \ve(z_k,\bz_k )\cw_{Q_k}( z_k, \bz_k )~,\ee
where  $z_k$ is the point on \cst\ to which $p_k$ asymptotes. 

We note that the operators $\cw_{Q_k}$ themselves do not create physical states. Rather the operators creating charged asymptotic particles are decomposed 
	into soft operators that transform nontrivially under large gauge transformations $\cw_{Q_k} $ and hard neutral operators $\widetilde O_{k} $ which decouple from the soft gauge field,
	\be \label{eq:decomp}
		O_k (p_k)  = \cw_{Q_k}(  p_k) \widetilde O_k (p_k)~.
	\ee
This decomposition into hard and soft factors is not unique, and our choice is unconventional. We define it here so that the soft part has strictly zero energy and transforms simply under \slc . This contrasts with Wilson lines localized to a line in spacetime which actually have infinite energy. 

Now we turn to the soft photon current $J_z$ which lives on \cst\ and, as  specified in the next section, is the difference of gauge field strength zero modes integrated along null generators of $\ci^\pm$  \cite{He:2014cra,Strominger:2017zoo}.   The soft photon theorem implies that large gauge transformations are generated by $\cs$-matrix insertions of contour integrals on \cst\ of 
$J_z$.  
 In the ultra-relativistic limit \eqref{ult}, 
\be \label{eq:jzphiOPE}
		J_z \cw_{Q_k}(z_k,\bz_k ) \sim\frac{Q_k}{z-z_k}\cw_{Q_k}(z_k, \bz_k ) ~.
	\ee
This is in turns implies the OPE 	
 \be \label{eq:jzphiOPE}
		J_z \phi^E (w , \bw) \sim- \frac{i}{z-w}~.
	\ee
It may be checked \cite{Campiglia:2015qka,Campiglia:2015lxa,Kapec:2015ena} that \eqref{eq:jzphiOPE} correctly reproduces the general large gauge transformation \eqref{nr}. 
	\section{Boundary gauge modes}
In this section we will express both the soft photon current $J_z$ and the electric Goldstone current defined as \be \label{sct} S^E_z = i\p_z \phi^E \ee 
 in terms of boundary values of the  photon field. A priori there are four independent  boundary values 
\be
		 A_z^{(0)} \big|_{\ci^+_+} ~,\quad  A_z^{(0)} \big|_{\ci^+_-} ~, \quad 
		 A_z^{(0)} \big|_{\ci^-_+} ~,  \quad  A_z^{(0)} \big|_{\ci^-_-} ~,
	\ee
	where $A_z^{(0)}$ is the leading, non-vanishing piece of the gauge field at $\ci$ and $\ci^+_\pm$ ($\ci^-_\pm$) are the future/past boundaries of future (past) null infinity.  
	The electric matching condition \cite{He:2014cra} equates two of these\footnote{This condition is modified in the presence of magnetic charges \cite{Strominger:2017zoo}.} 
\be\label{mtch} A_z^{(0)} \big|_{\ci^+_-}=
		 A_z^{(0)} \big|_{\ci^-_+} ~.\ee
 Another two combinations are related\footnote{In a gauge with $A_u\big|_{\ci^+}=0$ and $A_v \big|_{\ci^-} = 0$.} to ingoing and outgoing soft photon currents (see \cite{Strominger:2017zoo})
 \bea   e^2 \p_zN^-&=& \int dvF^{(0)}_{vz}= A_z^{(0)} \big|_{\ci^-_+}- A_z^{(0)} \big|_{\ci^-_-} ~,\cr e^2\p_zN&=& \int duF^{(0)}_{uz}=A_z^{(0)} \big|_{\ci^+_+} - A_z^{(0)} \big|_{\ci^+_-} ~.\eea
 The soft photon current is defined by time-ordered insertions of the difference
\be J_z=4\pi  \p_z(N^--N)=\frac{4 \pi}{e^2}\bigl(A_z^{(0)} \big|_{\ci^+_-}+A_z^{(0)} \big|_{\ci^-_+}- A_z^{(0)} \big|_{\ci^-_-} -A_z^{(0)} \big|_{\ci^+_+} \bigr).\ee
The equality (up to a sign) of the soft theorem for ingoing and outgoing
photons implies the vanishing of time-ordered insertions of\footnote{This invokes the standard assumption that the ingoing and outgoing vacua are the same. It is interesting to relax this but we shall not do so here. See \cite{kprs}.} 
\be\label{why} e^2(\p_zN+\p_zN^-)=A_z^{(0)} \big|_{\ci^+_+} - A_z^{(0)} \big|_{\ci^-_-} -(A_z^{(0)} \big|_{\ci^+_-}-A_z^{(0)} \big|_{\ci^-_+})=0~.\ee
A remaining linear  combination is  the Goldstone current 
	\be
		S_z ^E = \frac{i}{4 } \Big( A_z^{(0)} \big|_{\ci^+_+} +   A_z^{(0)} \big|_{\ci^+_-}+A_z^{(0)} \big|_{\ci^-_+} +   A_z^{(0)} \big|_{\ci^-_-} \Big)~.
	\ee
$S_z^E$ can be thought of as the `constant part' of $A_z^{(0)}$. 
Under a  large gauge transformation 		\be
		\delta_\ve S_z^E  = i\p_z{ \ve  }~,
	\ee
in agreement with \eqref{szw}. 	
	
	\section{Soft Wilson lines}
		\label{sec:SCETcurrent} 
		In this section, we complete the definition of the 2D Kac-Moody current algebra by specifying the OPEs of  the Goldstone currents and compute the soft matrix element of two Wilson operators.  Although the  physical picture is different, the computation of this section is reminiscent of that in \cite{Chien:2011wz}. \slc \ Lorentz-conformal invariance   dictates that all of the current OPEs are 
		proportional to  ${1 \over (z-w)^2}$.\footnote{Terms proportional to $\frac{1}{z-w}$ are absent because we are considering abelian gauge theory.} The soft photon current is itself neutral, so 
	\be \label{eq:jzjwOPE}
		J_z J_w \sim 0~.
	\ee
It follows from \eqref{eq:jzphiOPE} that
	\be \label{rdy} J_zS^E_w\sim{1\over (z-w)^2}~.\ee
Finally we have\footnote{The operators $\cw_{Q_k}$ introduced in the previous section are normal-ordered  so that such singular terms are removed   in the expansion of the exponent in powers of $\phi^E$. }
\be\label{lvl}			S^E_z S^E_w \sim \frac{k}{(z-w)^2}~,
		\ee
	for some constant $k$ which we now determine  by comparison with Wilson line expectation values. 
	
	First, we calculate the correlation function of two $\cw_{Q_k}$ operators.   We parameterize massive momenta as
	\be
			p_k^\mu = \frac{m_k}{2 \rho_k} \left (n^\mu + \rho_k^2 \hat {q}(z_k, \bz_k)\right)~,
	\ee
	with outgoing (incoming) asymptotic particles having $\rho_k>0$ ($\rho_k<0$).  The Wilson operators themselves are constructed from $\phi^E$ itself (rather than $S_z^E$) whose two-point function has ambiguities. To avoid these ambiguities, we differentiate and then fix the re-integration constants using \slc. Consider the following derivative of the logarithm of the two-point function 
	\be
		\begin{split}
			\p_{z_k}   \p_{z_\ell} & \log \langle \cw_{Q_k}(p_k)\cw_{Q_\ell}(p_\ell) \rangle \\
				  &= \left[i Q_k \int d^2z \p_{z_k} \cg(p_k; z, \bz)\right]
				   \left[i Q_\ell \int d^2w \p_{z_\ell} \cg(p_\ell; w, \bw)\right]\\& \quad \times 
				 	 \bigg[ \frac{\langle : \phi^E(z,\bz ) \cw_{Q_k} (p_k):: \phi^E(w,\bw ) \cw_{Q_\ell}(p_\ell) :\rangle}{\langle   \cw_{Q_k}(p_k)  \cw_{Q_\ell}(p_\ell)  \rangle}
					 		\\& \quad   \quad \quad \quad  \quad  
				- \frac{\langle : \phi^E(z ,\bz) \cw_{Q_k} (p_k): \cw_{Q_\ell} (p_\ell) \rangle\langle    \cw_{Q_k} (p_k): \phi^E(w ,\bw) \cw_{Q_\ell}(p_\ell) :\rangle}{\langle   \cw_{Q_k} (p_k) \cw_{Q_\ell} (p_\ell) \rangle^2} \bigg]~.
		\end{split}
	\ee
	Colons are used in the above expression to specify which products of operators have been normal-ordered (ie. singularities at zero separation have been removed).  The symmetry in the arguments of $\cg$ implies
	\be
		\p_{z_k} \cg(\rho_k, z_k, \bz_k; z, \bz) = - \p_z \cg(\rho_k, z_k, \bz_k; z, \bz) ~,
	\ee
and allows for the use of integration-by-parts to turn the scalar fields $\phi^E$ into their current counterparts $S_z^E$.  Then, using the operator product expansion of the $S^E_z$ currents, one finds
	\be
		\begin{split}
			\p_{z_k} \p_{z_\ell}  \log \langle \cw_{Q_k}(p_k)\cw_{Q_\ell}(p_\ell) \rangle
				 =    Q_k Q_\ell \int d^2z \int d^2w~ \cg(p_k; z, \bz)\cg(p_\ell; w, \bw)  \frac{k}{(z-w)^2}~.
		\end{split}
	\ee 
	 Next, we use the identity
	 \be
	 	\cg (\rho_k, z_k, \bz_k; z, \bz) = \frac{1}{2 \pi } \p_z \p_\bz \log \left ( p_k \cdot \hat q(z, \bz) \right) =   \frac{1}{2 \pi } \p_{z_k} \p_{\bz_k} \log \left ( p_k \cdot \hat q(z, \bz) \right) ~,
	 \ee
	 and integration-by-parts to evaluate the integral over the $(w,\bw)$ plane
	 \be  \label{eq:diffintermediate}
	 	\begin{split}
			\p_{z_k} \p_{z_\ell} &\log \langle \cw_{Q_k}(p_k)\cw_{Q_\ell}(p_\ell) \rangle 
				  =   -k Q_k Q_\ell \p_{z_k}\p_{z_\ell} \int \frac{d^2z}{2 \pi }\p_z \log\left ( p_k \cdot \hat q(z, \bz) \right)  \p_\bz \log\left ( p_\ell \cdot \hat q(z, \bz) \right)  ~.
		\end{split}
	 \ee
	Using the completeness relation
	\be
		\p_z \hat q^\mu \p_{\bz} \hat q^\nu  + \p_{\bz} \hat q^\mu  \p_z \hat q^\nu  = 2 \eta^{\mu\nu} + n^\mu \hat q^\nu + \hat{q}^\mu n^\nu~,
	\ee
	 the  remaining integral  may be rewritten as
	\be
		\begin{split}
			\int \frac{d^2z}{2 \pi }&\p_z \log\left ( p_k \cdot \hat q(z, \bz) \right)  \p_\bz \log\left ( p_\ell \cdot \hat q(z, \bz) \right) \\
				& \quad \quad =  \int \frac{d^2z}{2 \pi } 
				\bigg [\frac{p_k \cdot p_\ell}{\left(p_k \cdot \hat q(z, \bz) \right)\left(p_\ell \cdot \hat q(z, \bz) \right)}+ \frac{p_k \cdot n}{2p_k \cdot \hat q(z, \bz)}+\frac{p_\ell \cdot n}{2p_\ell \cdot \hat q(z, \bz)}\bigg]~.
		\end{split}
	\ee
	Upon integrating, the last two terms are $z_k$- and $z_\ell$-independent and consequently drop out of  \eqref{eq:diffintermediate}.  
	Moreover, the first term is \slc\ invariant on its own.  Hence after dropping the last two terms, we  may remove the derivatives on both sides to obtain 
 \be  \label{mediate}
	 	\begin{split}
			\log \langle \cw_{Q_k}(p_k)\cw_{Q_\ell}(p_\ell) \rangle 
				  =   -k Q_k Q_\ell  \int \frac{d^2z}{2 \pi } \frac{p_k \cdot p_\ell}{\left(p_k \cdot \hat q(z, \bz) \right)\left(p_\ell \cdot \hat q(z, \bz) \right)}  + c_0 ~,
		\end{split}
	 \ee
where $c_0$ is a soon-to-be fixed integration constant.
 The integral can be evaluated using Schwinger parameters, resulting in  
	\be\label{piu}
		\begin{split}
			 \int \frac{d^2z}{2 \pi } \frac{p_k \cdot p_\ell}{\left(p_k \cdot \hat q(z, \bz) \right)\left(p_\ell \cdot \hat q(z, \bz) \right)} 
				= - \gamma_{k \ell} \coth \gamma_{k \ell}~,
		\end{split}
	\ee
	where
		\be
			\cosh \gamma_{k \ell} \equiv \frac{p_k \cdot p_\ell}{\sqrt{-p_k^2} \sqrt{- p_\ell^2}}~. 
		\ee
This defines  $\gamma_{k \ell}$ up to shifts by $2\pi i$.
$\gamma_{k \ell}$ is real if one particle is outgoing and one particle is incoming, and vanishes if they are collinear (but not null). The result for two outgoing or two incoming particles, for which the righthand side is negative, is obtained by analytic continuation.  
		Namely,  defining a parameter $\beta_{k \ell}$ that is real for these configurations
		\be
			\cosh \beta_{k \ell} \equiv -  \frac{p_k \cdot p_\ell}{\sqrt{-p_k^2} \sqrt{- p_\ell^2}} ~,
		\ee 
		the desired result is obtained by replacing
		\be
			\gamma_{k \ell} \coth \gamma_{k \ell} \to  (\beta_{k \ell} - i \pi ) \coth \beta_{k \ell} ~
		\ee
		in \eqref{piu}.
	Focusing on the case of one outgoing and one incoming particle  and integrating \eqref{eq:diffintermediate}, one finds
		\be \label{erd}
			\langle \cw_{Q_k}(p_k) \cw_{Q_\ell}(p_\ell) \rangle =\delta_{0, Q_k+Q_\ell} \exp \big[k Q_k Q_\ell \left(\gamma_{k \ell} \coth \gamma_{k \ell} -1\right)\big]
				\equiv\delta_{0, Q_k+Q_\ell} \exp \left[-\Gamma_{k \ell} \right]~,
		\ee
		where $c_0$ was fixed by demanding that the two-point function reduce to unity when $Q_k + Q_\ell =0$ and $p_k+p_\ell = 0$ .

		The standard  quantum field theory result for the anomalous dimension of a pair of timelike Wilson lines is 
		\be\label{stn}
 		\Gamma_{k \ell} = \left \{\begin{array}{ll}
							-\Gamma_{\text{cusp}} Q_k Q_\ell\Big(\gamma_{k \ell} \coth \gamma_{k \ell} -1\Big)~, \quad & p_k \cdot p_\ell>0~\\
							- \Gamma_{\text{cusp}} Q_k Q_\ell \Big( (\beta_{k \ell} - i \pi ) \coth \beta_{k \ell}  -1\Big)~, \quad & p_k \cdot p_\ell <0~
							\end{array}\right.~,
		\ee  
	where the constant 
		\be\label{csp}
			\Gamma_{\text{cusp}} \equiv  \frac{e^2}{4 \pi ^2}~
		\ee
is known as the cusp 		anomalous dimension.  Exact agreement of our expression \eqref{erd} with this result follows if we identify Goldstone current algebra level \eqref{lvl} with 
\eqref{csp}
\be\label{csp2}
			k=\Gamma_{\text{cusp}} ~.
		\ee
This is a central result of this paper.

	This method for computing the correlation function of two Wilson operators will be  extended to the calculation of $n$-point functions of Wilson
		operators with both electric and magnetic charges in section 7.

 \section{Magnetic charges}
 	\label{sec:dualitycovariant}
 	
This section generalizes to theories with magnetically charged particles and non-zero  vacuum angle $\theta$.   
We focus on the ultra-boosted local operators of the form \eqref{ult}, but the results presented can be easily generalized to the more generic smeared operators associated to massive particles. 

The addition of magnetic charges is more than just a technical generalization. There is no known non-perturbative abelian gauge theory without magnetic monopoles, and so one might expect the full soft structure to reflect this. Indeed we shall find that the inclusion of magnetic degrees of freedom both complexifies and simplifies the soft structure relative to the purely electric case, leading to an elegant and simplified structure for the current algebra. 
\subsection{Duality covariant formulation}	 
	In the presence of  a $\theta$ angle 
	the allowed electric and magnetic charges $e_k$ and $g_k$ live on the charge lattice 
	\be e_k+ig_k=e(Q_k+\tau M_k),\ee
	where \be \label{eq:chargelattice}
		Q_k, M_k \in \mathbb{Z}~, \quad \quad \quad 
		\tau = \frac{\theta}{2 \pi }+ \frac{2 \pi i}{e^2}~ . 
	\ee
Magnetic charges lead to a 	 second set of  large gauge symmetries which can be interpreted as large  gauge transformations of the dual magnetic potential \cite{Strominger:2015bla}. These are naturally incorporated as a complexification of the original electric ones. The complexified gauge parameter lives on the torus defined by the  dual of the charge lattice 
\be\label{ltc} \ve\sim \ve + \frac{2 \pi i}{ \tau_2} \left (p+\tau q\right), \quad \quad \quad p,q \in \mathbb{Z}~.\ee

An important role in the following is played by the `shadow transform' which appears in CFT analyses\footnote{See \cite{SimmonsDuffin:2012uy}
for a recent discussion. The construction of a stress tensor from soft graviton modes in  \cite{Kapec:2016jld} also used a shadow transform.}. The shadow operator of a  dimension $(0,1)$ 2D operator, which we denote by a bar, is 
\be
		\bar{J}_z \equiv - \int \frac{d^2w}{2 \pi } \frac{1}{(z-w)^2}J_\bw~.
	\ee
	We will find it very convenient to discuss only $J_z$ and $\bar J_z$, and not $J_\bz$. No information is lost in this choice because  $J_\bz$ correlators can always be constructed from an inverse shadow transform using the identity
	\be
		\int d^2v ~ \frac{1}{2 \pi (z-v)^2}\frac{1}{2 \pi (\bw-\bar{v})^2}= \delta^2 (z-w)~.
	\ee

The gauge potential is not a duality-invariant concept and is not a priori well-defined in the presence of both electric and magnetic charges. However, the soft theorem involves a zero mode of the field strength and so is covariant. The soft theorem receives magnetic corrections 
	\be \label{eq:E&Msoft}
		\begin{split}
			\langle J_z O_1(p_1) \cdots O_n (p_n)\rangle & = \sum_{k = 1}^n \frac{Q_k + \tau M_k}{z- z_k} \langle  O_1(p_1) \cdots O_n (p_n)\rangle~, \\
			\langle \bar{J}_z O_1(p_1) \cdots O_n (p_n)\rangle & = \sum_{k = 1}^n \frac{Q_k +  \btau M_k}{z- z_k} \langle  O_1(p_1) \cdots O_n (p_n)\rangle ~,
		\end{split}
	\ee
	where $\bar{J}_z $ is the shadow transform of $J_\bz$.
In the absence of magnetic charges, the righthand sides are 
equal so we did not need to introduce $\bJ_z$. 
	
The soft sector has an electromagnetic duality symmetry even if the full theory does not.   Under $\text{SL}(2, \mathbb{Z})$		\be \label{eq:sl2z}
		\tau \to \frac{a \tau + b}{c \tau + d}~, \quad \quad \quad
		 \left (\begin{matrix} Q_k \\ M_k \end{matrix}\right) \to   \left (\begin{matrix}  a&- b\\- c&d\end{matrix}\right)  \left (\begin{matrix} Q_k \\ M_k \end{matrix}\right)~, \quad \quad \quad 
		  \left (\begin{matrix}  a&b\\c&d\end{matrix}\right) \in \text{SL}(2, \mathbb{Z})~.
	\ee
These relations  imply that the soft factor transforms as a modular form of weight minus one
	\be
		\frac{Q_k + \tau M_k}{z -z_k} \to  \frac{1}{c \tau + d}\frac{Q_k + \tau M_k}{z -z_k}~.
	\ee
	This in turn fixes the transformations of the soft photon currents 
	\be\label{dj}
		J_z \to  \frac{J_z}{c \tau + d}~, \quad \quad \quad \bar{J}_z \to \frac{\bar{J}_z}{c \btau + d}~.
	\ee
	 The Wilson operators \eqref{eq:defdressing} generalize to vertex operators of the form 
	 \be
	 	\cw_{(Q_k, M_k)}(z_k ) = \exp \left [\tfrac{i}{2} \left ( Q_k+ \tau M_k \right) \bar{\Phi}(z_k ) +\tfrac{i}{2} \left ( Q_k+ \btau M_k \right)  \Phi(z_k )  \right]~,  
	 \ee
	 where $\Phi$ is a complex chiral boson whose real and imaginary parts are the   electric and magnetic Goldstone bosons, $\phi^E$  and $\phi^M$, respectively. As befits a Goldstone boson, $\Phi$ lives on the same torus as   the gauge parameter	\be 
		\Phi \sim \Phi + \frac{2 \pi i}{ \tau_2} \left (p+\tau q\right)~, \quad \quad \quad \bar{\Phi} \sim \bar{\Phi} - \frac{2 \pi i}{ \tau_2} \left (p+  \bar{\tau}q\right)~, \quad \quad \quad p, q \in \mathbb{Z}~.
	\ee
Duality invariance of the vertex operators $\cw_{(Q_k, M_k)}$  fixes the transformations 	\be \label{eq:Phiduality1}
		\Phi \to (c \bar{\tau}+d) \Phi~, \quad \quad \quad \bar{\Phi} \to  (c \tau + d) \bar{\Phi}~.
	\ee
To reproduce the soft theorem, these scalars must obey the OPEs with the soft photon currents
	\be\label{pol} 
		\begin{split}
			\begin{array}{l l l}
		&J_z \Phi (w ) \sim 0 ~, \quad   \quad \quad \quad \quad
		& J_z \bar{\Phi}(w )\sim  -\frac{2i}{z-w}~, \\
		&\bar{J}_z \Phi (w ) \sim -\frac{2i}{z-w} ~, \quad \quad  \quad
		& \bar{J}_z \bar{\Phi}(w )\sim  0~.
		\end{array}
		\end{split}
	\ee
	It  follows that under large gauge transformations parametrized by a locally holomorphic function $\ve$
	\be
		\begin{split}
			\begin{array}{l l l}
		&\delta_\ve \Phi(z ) = 0~, \quad \quad \quad &\bar{\delta}_\ve \Phi(z) = 2 \ve(z )~,\\
		&\delta_\ve \bar{\Phi}(z ) = 2 \ve(z )~, \quad \quad \quad &\bar{\delta}_\ve \bar{\Phi}(z ) = 0~,
		\end{array}
		\end{split}
	\ee
	where $\delta_\ve$ and $\bar{\delta}_\ve$ are the transformations generated by contour integrals of $J_z$ and $\bJ_z$, respectively. 
	As before, Goldstone currents can be constructed from the Goldstone bosons
	\be
		S_z = i \p_z \Phi~, \quad \quad  \quad \quad \quad \bar{S}_z  = i \p_z \bar{\Phi}~,
	\ee
	where these obey	\be\label{ds}	
		S_z \to  (c \bar{\tau}+d)S_z~, \quad \quad \quad \bar{S}_z \to  (c\tau+d) \bar{S}_z~.
	\ee
Demanding the desired result 	\eqref{stn} for purely electric Wilson operators together with duality invariance implies	\be
		S_z S_w \sim 0~, \quad \quad \quad \bar{S}_z \bar{S}_w \sim 0~, \quad \quad \quad S_z \bar{S}_w \sim \frac{2\Gamma_{\text{cusp}}}{(z-w)^2}~.
	\ee

\subsection{Current identifications}	
	It might seem from our treatment so far that $S_z$ and $J_z$ are independent 2D currents. However in this section we argue that in fact there is a redundancy and one may consistently set 
	\be\label{id} S_z=\gc J_z~, \quad \quad \bar S_z=\gc \bar J_z ~,\ee
reducing the current algebra to  a single complex current.    \eqref{id} is consistent with the duality transformations \eqref{dj} and \eqref{ds} as well as the OPEs \eqref{pol}. Without such an identification, we have too many fields in the sense that soft photon insertions are described by $J_z$, and $S_z$ do not comprise a second set. $S_z$ was introduced as the Goldstone mode and appears in   Wilson operators associated to charged particles. The identification \eqref{id} asserts that the $J_z,\bJ_z$ current algebra alone can compute the full soft factor of the $\cs$-matrix. 
	
	The identification \eqref{id} immediately implies the OPE
\be\label{jbj} J_z \bar J_w \sim {2 \over \gc (z-w)^2}.\ee
One might expect that this relation could be unambiguously checked from the definition of 
the soft photon current as the soft $\omega \to 0$ limit of  the photon field operator. However, it turns out that the OPE of $J_z$ and $J_\bw$ has a $\delta^2(z-w)$ contact term proportional to $\omega \delta (\omega)$. The latter depends on the detailed manner in which $\omega \to 0$ and can be defined to take any value.  This is familiar: it often occurs in  CFT that a prescription is needed for contact terms. Constructing $\bJ_z$ as the shadow transform of $J_\bz$, the ambiguity in the contact term becomes an ambiguity in the   ${1 \over (z-w)^2}$ term in the 
$J_z\bJ_w$ OPE. Hence at this level \eqref{jbj} can simply be regarded as a prescription for the contact term.

Given this prescription for the contact term,  we define the two currents 
\be B_z=S_z-\gc J_z~,~~~~\bar B_z =\bar S_z-\gc \bar J_z~.\ee  It is then easy to check that all OPEs involving 
$B_z$ or $\bar B_z$ vanish:
\be\label{zro}B_z \bS_w\sim B_z \bJ_w \sim \bar B_zS_w\sim\bar B_z J_w \sim 0~.\ee
 Hence it decouples and may be ignored. 
 This suggest that we  quotient out the $B_z$ current algebra  and view $J_z$ as the fundamental current with OPE \eqref{jbj}. This current is related to the Goldstone mode by 
\be J_z={ i\over \gc}\p_z \Phi~, ~~~~~ \bJ_z={ i\over \gc}\p_z \bar \Phi~.\ee

 	 \subsection{Sugawara stress-energy and central charge}
	 
	 	The Sugawara stress-energy tensor is obtained by inverting the matrix of current levels
		\be
			T_{zz} =  \frac{1}{2} \Gamma_{\text{cusp}} J_z \bJ_z  ~.
		\ee
Insertions (or scattering of) this operator generate the full infinite-dimensional group of local conformal transformations (not just \slc) on the currents. 		
The conformal  weight (or dimension) of a vertex operator $\cw_{(Q_k, M_k)}$ is 
		\be
			h =\frac{\Gamma_{\text{cusp}}}{2}\left |Q_k + \tau M_k \right|^2 = \frac{ \left |Q_k + \tau M_k \right|^2}{4 \pi \tau_2}~.
		\ee
It would be interesting to explore the possibility of enhanced symmetries at values of the coupling for which this becomes integral. 		To compute the central charge, we note the OPE		\be
			T_{zz} T_{ww}   \sim \frac{1}{  (z-w)^4}+ \cdots ~,
		\ee
giving a central charge 
		\be
			c = 2~.
		\ee
		\section{Non-perturbative soft $\cs$-matrix}     
	Soft factors associated to generic scattering processes appear in our language as (connected)\footnote{The product of $J_z$'s and $\bar J_z$'s has been normal-ordered to remove contact terms arising from
	 the non-zero 
	current algebra level \eqref{jbj}, since these correspond to disconnected Feynman diagrams.} correlation functions of soft photon currents and Wilson operators
	  \be   
      \langle  J_{z_1}   \cdots J_{z_r} \bJ_{z_{r+1}}  \cdots \bJ_{z_m}   \cw_{(Q_{ 1},M_{1})} \cdots \cw_{(Q_{n },M_n)}  \rangle ~. 
      \ee  	 
     To evaluate these correlation functions,   the soft photon current algebra is first used to evaluate the contributions from insertions of $J_z$ and $\bar J_z$ and obtain an expression only involving 
     a correlation function of $\cw_{(Q_k,M_k)}$ operators
      \be \label{eq:softhm}
      	\begin{split}
      \langle  J_{z_1} & \cdots J_{z_r} \bJ_{z_{r+1}}  \cdots \bJ_{z_m}   \cw_{(Q_{ 1},M_{1})} \cdots \cw_{(Q_{n },M_n)}  \rangle\\&
      =  \prod_{i = 1}^r \prod_{k = r+ 1}^m   \bigg[\sum_{j =  1}^{ n} (Q_j+ \tau M_j) \p_{z_i} \log \big( p_j \cdot \hat {q} (z_i, \bz_i) \big)  \bigg] 
	  \bigg[\sum_{\ell=  1}^{ n} (Q_\ell+ \btau M_\ell)\p_{z_k} \log \big( p_\ell \cdot \hat {q} (z_k, \bz_k) \big)   \bigg]  \\& \quad \quad     \times 
	  \langle     \cw_{(Q_{ 1},M_{1})} \cdots \cw_{(Q_{n },M_n)}  \rangle~.
	 	\end{split}
      \ee  
      
      Then, the correlation function of $\cw_{(Q_k,M_k)}$ operators can be calculated by applying the same technique used in section 5
      \be  \label{eq:softmatrix}
      	\begin{split}
     \langle  J_{z_1} & \cdots J_{z_r} \bJ_{z_{r+1}}  \cdots \bJ_{z_m}   \cw_{(Q_{ 1},M_{1})} \cdots \cw_{(Q_{n },M_n)}  \rangle\\&
      =  \prod_{i = 1}^r \prod_{k = r+ 1}^m   \bigg[\sum_{j =  1}^{ n} (Q_j+ \tau M_j) \p_{z_i} \log \big( p_j \cdot \hat {q} (z_i, \bz_i) \big)  \bigg] 
	  \bigg[\sum_{\ell=  1}^{ n} (Q_\ell+ \btau M_\ell)\p_{z_k} \log \big( p_\ell \cdot \hat {q} (z_k, \bz_k) \big)   \bigg]  \\& \quad \quad     \times 
      \delta_{0,\sum_s Q_s} \delta_{0,\sum_t M_t} \prod_{\substack{p,q=  1\\p \neq q}}^{n}\exp \left(- \Gamma_{pq}\right)  ~,
      	\end{split}
      \ee
      where $\Gamma_{k \ell}$ now takes the duality invariant form
     \be
			\Gamma_{k \ell} = \left \{\begin{array}{ll}
							- \frac{e^2 \vec Q_k \cdot \vec Q_\ell }{4 \pi^2} \Big(\gamma_{k \ell} \coth \gamma_{k \ell} -1\Big)~, \quad & p_k \cdot p_\ell>0~\\
							- \frac{e^2 \vec Q_k \cdot \vec Q_\ell }{4 \pi^2} \Big( (\beta_{k \ell} - i \pi ) \coth \beta_{k \ell}  -1\Big)~, \quad & p_k \cdot p_\ell <0~
							\end{array}\right.~,
	\ee  
	with
	\be\label{dew}
		\vec Q_k \cdot \vec Q_\ell  \equiv \frac{1}{2}  \left (Q_k + \tau M_k\right)\left (Q_\ell + \btau M_\ell\right)+\frac{1}{2}  \left (Q_k + \btau M_k\right)\left (Q_\ell + \tau M_\ell\right).
	\ee    	Setting all magnetic charges $M_k$ to zero, one recovers the result for electric large gauge symmetry. 	 
	On the other hand, when the asymptotic particles carry magnetic charge, we have derived a new and fully $\text{SL}(2, \mathbb{Z})$ duality covariant expression for the soft factor associated to the complexified large gauge symmetry
	of Abelian gauge theory.  We conjecture that \eqref{eq:softmatrix} is nonperturbatively exact in theories with both electric and magnetic charges.

		\section{Soft factorization}
		
		\label{sec:softfactorization} 
		
The soft factorization theorem in  QED   states that, to all orders in perturbation theory, the  $\cs$-matrix 
factorizes into a hard and a soft part, where all collinear and soft poles reside in the soft part, and the hard part is finite. 
Precise statements may  be found in \cite{Bassetto:1984ik,Feige:2014,SchwartzQFT} and references therein. Roughly speaking  the scattering amplitudes obey     \be \label{eq:QEDsoft}
      		\begin{split}
			\langle \cdots q_{m}; \cdots k_{r} ; \cdots p_{n}\vert \cs   \vert q_{1}\cdots;  k_{1} \cdots ; p_{1} \cdots \rangle 
				\overset{q_i\text{ soft}}{\longrightarrow}& \\
					\underbrace{\langle \cdots k_{r} ; \cdots \hat p_{n}  \vert   k_{1} \cdots ;\hat p_{1} \cdots \rangle   }_{\text{hard}}&
					\underbrace{ \langle \cdots q_{m}  \vert  \cw_{Q_n}(p^n)  \cdots \cw_{Q_1}(p^1) \vert q_{1}\cdots \rangle       }_{\text{soft}}~.
		\end{split}
      \ee
      In the above formula, $q_i$ ($k_i$) represent asymptotic single-particle soft (hard) photons
      and  $p_j$ represent asymptotic hard electrons (or positrons). The field which creates asymptotic hard electrons $\psi(p_k)$ is decomposed as $\psi(p_k)\sim\hat\psi(p_k)\cw_{Q_k}(p_k)$ and $\hat p_k$ 
      in \eqref{eq:QEDsoft} denotes a `stripped' asymptotic electron which decouples from the soft photon. The factorization occurs 
      at leading order in a physical power counting parameter $\lambda$ that sets the soft scale, thereby distinguishing  soft from hard momenta.
        While there is not a unique choice for this IR scale, it is generally defined by $\frac{q_i^0}{p_j^0} \sim O(\lambda)$ where $q_i^0$ and $p_j^0$ are the energies of the 
       soft and hard asymptotic states respectively.
       
       The full amplitude on the left hand side of \eqref{eq:QEDsoft}  diverges as $ {1 \over \Omega}=\prod_{i=1}^m (q_i^0)^{-1}$ as $q_i\to 0$.\footnote{There are additional soft divergences due to virtual photons, which
       are the leading divergences in the absence of external soft photons $(m=0)$.}
        These are the soft poles, and their residues may be  obtained simply by multiplying by $\Omega$ 
          and setting $q_i=0$.\footnote{There is also universality in the subleading term in the $q_i$ expansion that would be of interest to study.}  The structure of the resulting leading soft amplitude  
          is reproduced by  \eqref{eq:softmatrix}-\eqref{dew} specialized of course to electric charges only. 
          In particular, the singular terms in the OPE between soft photon currents and Wilson operators reproduce the leading-order tree-level exact soft factor \cite{Bassetto:1984ik,Feige:2014,SchwartzQFT}.
          In addition, the correlators of Wilson operators reproduce the exponentiated leading-order 1-loop exact virtual soft divergence, where the IR divergent prefactor is removed by our procedure of matching the
          anomalous dimensions of 2D and 4D operators.
          We conclude that the 2D $J_z,\bJ_z$ current algebra computes the soft amplitude of the QED $\cs$-matrix.  Rewriting the $\cs$-matrix as a correlator on \cst\ and using the hard/soft decomposition defined in \eqref{eq:decomp}, the soft factorization theorem is 
       simply the familiar 2D current algebra factorization formula
      \be\label{mds}
      \begin{split}
      	\langle J_{z_1} \cdots J_{z_{r}} \bJ_{z_{r+1}}& \cdots \bJ_{z_m}  O_{ 1} \cdots O_{n }   \rangle \\
      &	= \underbrace{\langle  \widetilde{ O}_{ 1} \cdots \widetilde{O}_{n }  \rangle}_{\text{hard}}
		\underbrace{\langle  J_{z_1} \cdots J_{z_{r}} \bJ_{z_{r+1}} \cdots \bJ_{z_m}   \cw_{(Q_{ 1},M_{1})} \cdots \cw_{(Q_{n },M_n)}  \rangle}_{\text{soft}} 
      	 ~,
      \end{split}
      \ee 
      specialized to electric charges only.
      
      In conclusion, the soft factorization theorem in abelian gauge theory 
      and current algebra factorization of 2D CFT correlators are the same thing. Moreover, 
   2D  current algebra techniques  are computationally effective for determining the soft $\cs$-matrix of 4D abelian gauge theory.

\section*{Acknowledgements}
The authors are  grateful  to D. Kapec, P. Mitra, S. Pasterski, S.-H. Shao and especially  M. Schwartz for useful conversations. This work is supported by DOE grant DE-SC0007870.

\end{document}